# Recordism: A social-scientific prospect of blockchain from social, legal, financial, and technical perspectives

Zihao Li, Hao Xu*, Yang Fang, Boyuan Zhao, Yinuo Liu and Lei Zhang, *Senior Member, IEEE*

*Abstract*— Blockchain technology has the potential to revolutionize the architecture of cyberspace by transforming the way information is stored, circulated, and exchanged in cyberspace through decentralization, transparency, and de-identification. This means that ordinary participants can simultaneously become traders, miners, retailers, and customers, thus breaking down barriers, reducing the information gap between participants in the community, and contributing to the futuristic metaverse with an open, progressive, and equal ideology. The impact of this information transformation empowered by blockchain extends to our understanding of methodology, legal governance in cyberspace, and financial and technological development. This study asks: what are the implications of the blockchain-driven information revolution for society and social sciences? In order to answer this main question, the paper focuses on four key perspectives: methodological, legal, financial, and technical. Through the analysis of these four perspectives, the paper provides a comprehensive understanding of the impact of blockchain on society, the social sciences, and technology, making a contribution to current scholarship. It finds that blockchain is not only an innovative cognition method, but also a community representative, serving as a source of trust, a governance watchdog, an enforcer of cyber laws, and an incubator for future technologies. Despite some challenges in integrating blockchain with existing social structures, this paper concludes that blockchain has the potential to play a significant role in shaping the future.

## I. Introduction

The past few years have ushered in an enormous shift towards digitization in our daily lives, especially with the widespread impact of COVID-19. In the context of less in-person interaction and the pandemic panic, trust among individuals and society have become more fragile than ever [1]. However, the advent of Distributed Ledger Technologies (DLTs) [2], a.k.a. blockchain, provides us with a cryptography-based trust mechanism. Due to its unique framework and mechanism, blockchain reforms cyberspace and transforms the storage, circulation, and exchange of information through decentralization, transparency, and pseudonymization [3]. These features deeply influence society, moving it from a centralized social world to a brave decentralized world. Such a decentralization mechanism through technology shakes the very foundation of society, i.e., trust, as all the trust is secured by every participant rather than any authority we know from reality. Therefore, the information transformation empowered by blockchain also has a profound impact on our methodological cognition, legal governance in cyberspace, and financial and technological development [4].

The first shift enabled by blockchain is methodological cognition. Blockchain-driven information revolution refreshes the cognition of the source of governance in society and challenges the traditional centralized source of governance. In contrast, a blockchain-based framework provides a decentralized source of governance where anyone could be a part of the source to establish trust and governance. More importantly, from a methodological perspective, blockchain reshapes the world with its design philosophy. It can act as a panoptical instrument to boost the self-discipline of individuals. On the other hand, the governor is also restrained from abusing their power, as all actions are recorded on the chain and are transparent to everyone [5]–[7]. Therefore, the blockchain-driven information revolution realigns the balance between individuals and the government. It may also change society, as the source of power is ultimately bottom-up and there is no information asymmetry. This paper argues that such information transformation could lead to a new methodological cognition called "recordism" (a term for the prescribed idea of attributing the significance of recording elements of individuals and their lives).

Secondly, the blockchain-enabled information revolution has great influence on the legal governance of cyberspace [5], [6]. Because of the decentralization of the blockchain framework, trust among individuals and cyberspace is re-established through cryptography, which also alters the foundation of legislation [8]. Accordingly, there has been a corresponding shift in legal regulation (from a single regulatory body to multiple regulatory bodies) and changes in the approaches and methods to legal research (code as law). Furthermore, blockchain refreshes many legal notions (e.g., those of trust, transparency, etc.), the forms of legislation (e.g., Machine-Readable legislation), and alters the form of electronic evidence and legal executions (e.g., the legal execution issues of data protection law) [9]. Also, if viewed as instrumentality from a philosophical perspective, a blockchain-based system provides us with a new view of the law: that we make law as *de-facto* the product of consensus and keep it to ourselves, independently. To further experience the magic of the blockchain-law relationship, we have a vision that native blockchain law lives longer than the community that made it, especially in the private sector, as it is maintained and modified by contemporary groups of stakeholders - long live the blockchain native law!

Thirdly, the blockchain-enabled information revolution also enables many new applications in the business and financial area [2], [3]. As the Bitcoin market, the forerunner of the blockchain application, mined its genesis block in 2008[10], and has continued to grow and thrive, it has transformed the finance market and challenged the old world of centralization. However, Bitcoin does not unleash the full power of blockchain as blockchain means much more than just cryptocurrency [3], which was already an earth-breaking application based on the decentralization idea. Meanwhile, in the areas of business and finance, blockchain brings more revolutionary changes [7]. For example, smart contracts powered by blockchain could improve the efficiency of the financial market and optimize corporate governance. Also, various central banks establishing their own digital currencies is another example of the influence of blockchain on the financial sector [11].

As well as the abovementioned impacts of blockchain on society, we expect to see blockchain deepening its impact on our day-to-day business, the way we think, and the principles we trust. In fact, philosophy has predicted a world with tools of similar functionality in Michel Foucault's work [12], namely the panopticon[1]. The theoretical basis of the present paper is Foucault's panopticism, which argues that the power dynamics in modern society are shaped by centralized surveillance and control mechanisms aimed at achieving a communal goal. The transparency enabled by blockchain technology highlights the power of being seen, where all the participants in the system can share the benefits of being part of the "panopticon tower" even though no one is actually watching them. This leads to an ethical transformation of pseudo-anonymous activities and to reflection on the structure of power in decentralized systems and governance. Now that the tool has finally become available to the general public, we are excited to describe the instrumentality of blockchain and how it widens our path to power and self-discipline.

Blockchain not only has a great societal impact, but also demonstrates huge potential in combination with other technologies if it is regarded as an innovative methodological cognition [13]. The combination of blockchain and other technologies can unleash the greater potential of a diverse range of other technologies and blockchain itself. For example, combining functional consensus and blockchain could make blockchain more environmentally-friendly, and the extra thermal energy generated by the blockchain could feed into room heating systems. Also, the conjecture of a combination of a revolutionary storage medium (e.g., biological genetic materials, including DNA and RNA) and blockchain could provide new solutions to distributed storage and data handling problems.

In the early stages of blockchain, Aste et al. predicted its societal impact and concluded that blockchain has groundbreaking potential through disintermediation, cryptography, and immutability, enabling a new decentralized peer-to-peer market [14]. Although Aste et al. acknowledged that blockchain has some weaknesses, such as efficiency and physical limits, the existing literature does not investigate blockchain's societal implications in detail [14]. In response to Aste et al.'s conclusion, this paper therefore focuses on two main research questions: **(i) What are the implications of the blockchain-driven information revolution for the social sciences? (ii) How can the impact be positively harnessed by society and social sciences?** In order to answer these two research questions, this paper's analysis is from four perspectives: methodological, legal, financial, and technical. These four perspectives will provide a more comprehensive analysis of blockchain's impact on society and the social sciences to contribute to the current scholarship.

The paper is structured as follows: a brief introduction to, and definition of, blockchain and its features that could greatly impact society and the social sciences is presented. Then, the paper gives details of how blockchain as a novel methodology and instrumentality transforms the concept of the source of power, discipline, and information inequality. Following that, the legal implications of the blockchain-driven information revolution will be discussed in relation to theoretical (e.g., the basic notions of trust and transparency) and practical (electronic evidence, privacy, and data protection) impacts. In addition to law, blockchain's financial and economic envisioning is discussed, ranging from corporate governance to central banks and digital currencies. Lastly, the paper analyzes the potential of blockchain-enabled scenarios with other technology, from functional consensus to futuristic biological genetic material, and how the technology responds to challenges raised in the legal and financial industries.

II. RESEARCH METHOD

This paper mainly presents desk-based research exploring the potential implications of blockchain on society, legislation, and the financial and technology sectors. The analysis was conducted using the legal doctrine method (so-called 'black-letter' approach) from a socio-legal perspective. The legal doctrine method treats case-law and statutory measures as primary sources, while background policy papers, academic commentary, and analysis are used as secondary sources. The main focus of legal doctrine method is interpreting the law with reference to precedents in the context of blockchain. Through this analysis, the implications of blockchain on law can be presented and potential gaps can be identified. An interpretive analysis is not limited to legislation, but extends to policy reports, authority guidelines, and academic articles. From a socio-legal perspective, societal impact and financial impact are discussed. This paper reviews several essential industrial reports and seminal literature, reflecting on the implications of blockchain for society and the financial sector. The implications and issues extracted from legal, financial, and societal perspectives serve as the basis to enable the technological discussion, which responds to the identified issues by proposing interdisciplinary solutions.

III. WHAT IS THE BLOCKCHAIN?

The term 'blockchain' refers to an old concept of chaining data together which was proposed in the mid-1980s [7]. It has now become a pillar of the booming cryptocurrency technology, as it enables a trustless, decentralized, and tamper-proof ledger [10]. As its name suggests, the data are stored in the form of blocks [10], chained together by the previous block's hash, as illustrated in Figure 1, to ensure data integrity and security. Hence the later block can locate itself from the current block's hash value, which is a unique and irreversible identifier based on the content in the block. In most cases, this manner of distribution and the consistency of blockchain is decided by consensus protocols, a.k.a. consensus algorithms, by which data processing and storage formation is embedded when each node runs.

---

[1] The panopticon is a theorized type of institutional building and system of control designed by the English philosopher and social theorist Jeremy Bentham in the 18th century. The concept of the design is to allow all prisoners of an institution to be observed by a single security guard, without the prisoners being able to tell whether or not they are being watched.

Blockchain networks are often described as decentralized networks with countless benefits based on their autonomous and self-sovereign nature. Decentralization is achieved in two ways – first, the decentralization of administration is done by the consensus mechanism that sets up the ground rules of the crowdsourced network. The network utilizes the distributed consensus among participants to reach a mutual decision for every action on the network. Distributed consensuses can be roughly grouped into Proof-based (PoX) consensus, which selects the miner by competing in solving a hard problem; or Voting-based (e.g., RAFT and PBFT), which uses the identity of the participants to form a quorum to reach fault-free decisions (see Figure 1). Second, blockchain is built upon a Peer-to-Peer network, which makes use of decentralized routing protocols and content delivery mechanisms. With the help of the P2P network, the blockchain keeps a copy of the ledger in every participating node, making the blockchain a truly decentralized database based upon decentralized networks.

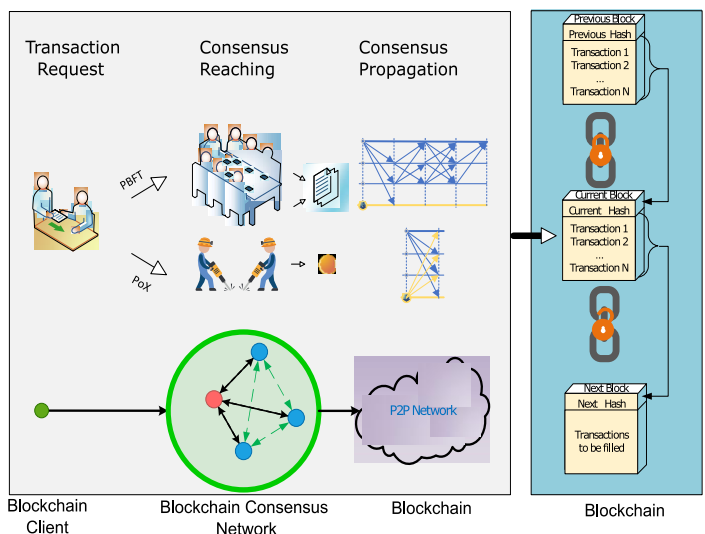

Figure 1: How blockchain works

Unlike the traditional centralized system, trust in the blockchain system is established through cryptography [10], and the blockchain can be roughly divided into public chains and private chains based on their access settings. In a public chain, anyone is free to join and participate in the core activities of the blockchain network. Therefore, in a public chain, blockchain allows any user to store a local copy of the blockchain and to propose new blocks, which lay down the foundation for a trustless environment.

Another appealing aspect of blockchain technology is the degree of transparency that it can provide [7]. Based on its design philosophy, blockchain is a transparent framework in which anyone can join the network (especially in the public chain), and therefore, everyone can see all the information on the chain. In the Distributed Ledger Technologies (DLT) mechanism, data is grouped into each block and bonded with the existing blocks through hash check and consensus algorithm. The data on the chain include different metadata; for instance, the hash of all transactions contained in the block, a timestamp, and a hash of the previous block. In terms of transparency, participants are able to read the distributed ledger, and see the changes to the ledger, hence there is a total transparency of transactions. In fact, the promised transparency does not reveal user identity thanks to the anonymous and non-linkable identities used by the clients. Hence, the transparency is achieved across all transactions and records, where all details of exchanged information are known, without knowledge of who the parties involved are. User identities are never revealed in the network unless the user declares it. The hash-based cryptographic chain employed in this system ensures that all records exhibit traceable and append-only attributes, thereby enhancing transparency and accountability. Through its information traceability, blockchain can safeguard transparency as nothing is unduly modified without recording the changes, and as a result, everything is transparent, except for user identities.

Additionally, blockchain can enable self-tamper-proof and self-authentication applications [7], [10]. As every block is linked and verified through the hash algorithm, data integrity and security can be guaranteed. If necessary, the blockchain network can trace back to a historic transaction without any third parties. Therefore, blockchain-based applications can be self-tamper-proof and achieve self-authentication.

IV. BLOCKCHAIN AND THE SOURCE OF GOVERNANCE

Blockchain has shaped the recording of humanity, creating a new world where no records can easily be altered, erased, or disguised by preserving every single transaction by blockchain participants. In the traditional thinking on power, information asymmetry is one of the most important sources of power [15], and people are classified based on their disclosure of information. The blockchain-driven information revolution has initiated a permanent transformation, as its inherent transparency offers boundless opportunities for ordinary participants to gain a comprehensive understanding of society, devoid of gaps or asymmetry. This advancement transcends the paper-based medium and even surpasses the scope of the Internet, where information remains in the hands of a select few users [16]. The origin of the panoptic tower, which was first proposed as a prison model [12], [17], provides analytical thinking on a utopia in which everyone is watched by a fantastical tower that knows all the members' secrets, which in fact is not far away from daily practice in a surveillance state. The blockchain can also regulate individuals, just like any centralized government. The panoptic tower is represented by the governors of blockchain, which can see through the identities of blockchain network users and enforce orders based on their data. The blockchain has therefore been given an instrumental power in philosophical terms.

The panoptic model can be applied to reality, essentially helping to develop a sophisticated, technically proven, and cryptographically safe panoptical instrument with trustable records, identity pseudonymization, and transparency for everyone without tempering the prized human rights of privacy and freedom.

A panoptical instrument for personal behaviors is a tool allowing humans to reach a society with better order, discipline, honesty, and reality. During the process of recording more behaviors on the blockchain - for instance, daily transactions and interactions with civil services, commercial transactions, jury records, lawsuits, healthcare records, etc. - the value carried by blockchain increases, and the enforcement of social order can be greatly eased. The transparency of personal data in certain fields related to the governor provides an opportunity to enrich the conceptual scope of surveillance [15], discipline, and punishment [12]. Meanwhile, because they are monitored by an authority, individuals are motivated to self-discipline and behave responsively towards society and the community, creating the new reality of a transparent society.

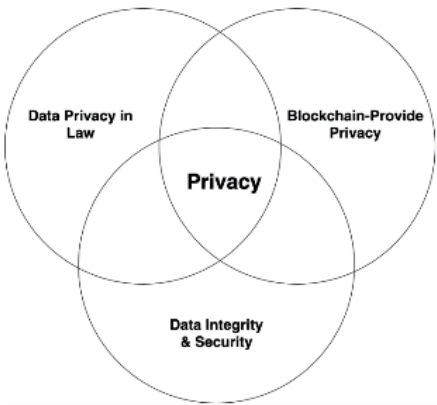

Figure 2 Intersections of blockchain impacts

A. Privacy

Some scholars argue that blockchain's mechanism may undermine individuals' privacy as personal data is shared and users are 'gazed' upon by others [18], [19]. Furthermore, blockchain decentralizes users' personal data into various devices, which not only causes difficulties in legal enforcement, but is also not compliant with the data minimization principle in data privacy law [18], [19]. Therefore, blockchain-based infrastructures compel us to rethink the obsolete design models of privacy and to have the ability to establish a new standard of encrypted privacy with data integrity, security, and transparency. To some extent, blockchain-based infrastructures could further the achievement of legal objectives. Firstly, user identity and other identification data is secured through cryptography, which guarantees that a user's identity is not disclosed to any third party. Secondly, the tension between the law and blockchain (because the law has always lagged behind the improvement of technology) can be reconciled. **Error! Reference source not found.** shows that data privacy law and blockchain share common objectives: to provide users with more feasibility and accessibility to protect their personal data. Although data privacy law and blockchain use different methods to protect users' privacy, combining both could establish better data privacy protection and renew the existing data protection architecture. As **Error! Reference source not found.** demonstrates, one of the goals of data privacy law is to protect data subjects through the principle of data integrity and transparency, which forms modern data protection law, e.g., the General Data Protection Regulation (GDPR) [20]. DLTs can partly achieve the objectives of data privacy law through Privacy-by-Design (PbD), by which the decentralized framework and zero-trust mechanism[2] [21] can guarantee the integrity of personal data and the transparency of the network. Unlike the right-based legal system, blockchain ensures user privacy through a cryptography-based mechanism. Although in a public chain[3] everyone can see the entire information on the chain to verify a transaction, all data is encrypted by asymmetric cryptography. The zero-trust mechanism enables users to access the information without any intermediary. Meanwhile, a blockchain-based mechanism also allows users to control access to their personal data by third parties on a case-by-case basis. Thus, implementation of the right to access is strengthened in practice. Therefore, the blockchain brings three advantages to data privacy: (1) it secures the uploaded data; (2) it offers a decentralized mechanism to avoid data breaches or an oligarchy; and (3) it records all data accessing and activity to provide users full oversight of their data.

B. The Shift of Governance

From another perspective, blockchain protects the user against the government. Because the governor is disciplined by all the transparent records accumulated in history, they may refrain from touching any individuals in the regime of blockchain social practice, as any actions against individuals will be seen on the blockchain by all communities. Therefore, a balance is formed between individuals and governments. No one wants to leave evidence of inappropriate records nor to weaken the power by leaving bad records on actions to individuals. The longer the record exists in the community, reputations and deterrence will become as strong as the record itself. This opens up an opportunity if the governance originates from the record rather than the political quorum of representatives. Whoever vows to provide the community with civil service shall commit to protecting records, as this is the source of power without any political interests.

The shift of power from the governing body to the records held by blockchain transforms the whole of society. It reforms the community from the bottom to the top and reduces the information gap between users [5], which later will result in a reduced class gap. A new ethics emerges when the recordism populates the space and time of the society, leading to a new scale of power concentration, as there is no reason to stop concentrating the power if the power is also gazed upon by the public and recorded in a publicly visible medium, i.e., the blockchain. This concentration is a revelation of the new democracy that can settle disputes in no time.

Thanks to the nature of the distributed ledger, blockchain provides a new scope of power regulation. By recording the power in action, blockchain improvises the power in both the past and the future (by auditing past actions and supervising future legislation). The power also benefits from the records of all individuals who are self-regulating themselves thanks to the transparency of the records. No comparable feature in history has provided such functionality bound to the record in relation to every individual and every action. The theory of recordism is becoming the principle of society in the near future, no matter how the adoption of blockchain is realized. In fact, the modern world has long been an information-based society, but blockchain provides a clearer path to freedom and democracy where privacy, trust, and transparency can be preserved by design.

---

[2] Unlike the traditional approach based on the assumption that a third-party organization is trustworthy, every service in the DLTs is validated without selective trust. DLTs upholds the "verify and never trust" principle to provide enhanced security and integrity.

[3] A public blockchain is a decentralized platform accessible by anyone. In other words, it is permissionless, meaning that anyone can join, and write and read the information on the platform.

Running a government on blockchain may initially sound impossible for both technical and societal reasons. First, the blockchain suffers from limited overall throughput, which limits the concurrent users of the blockchain network. This issue raises an inconsistency of administration when the data are frequently modified, and require assistance from asynchronous data to speed up the administration, leaving integrity behind. On the other hand, administration record needs to be stored for a number of years (or permanently) to serve its purpose, bringing a huge load of storage as all records are distributed equally into every full node of the blockchain. Such storage may be so accumulated in time that it becomes too heavy to mitigate. Moreover, blockchain-based governance may raise social barriers for the public to learn and be aware of the source of governance. It requires new skills training and education in order to increase public knowledge, thus posing a new societal threat.

## V. The Impact of Blockchain on Law

The advent of blockchain alters the cyber framework and transforms information storage, circulation, and exchange through decentralization, transparency, and pseudonymization. Data controllers and processors can be decentralized and maintained through consensus algorithms to guarantee integrity and transparency [22]. User's identities are protected through multiple tiers of encryption to achieve pseudonymization. This transformation accordingly brings new opportunities and challenges as to how the law normally regulates online information-related areas. This section will discuss the key question of the impact of the blockchain-enabled information revolution on the law. It will analyze the theoretical impact of blockchain on law, including cyber law theory, the foundation of trust, and online transparency from a data protection perspective. From a practical perspective, it analyses blockchain-based machine-readable legislation, the impact of blockchain-based electronic evidence systems and, most importantly, the influence of blockchain on data protection law. The law with autonomous execution capability and the transparency of code as law are the critical enablers of blockchain practice in legislation, where legal regulators and the public each benefit from the trustworthy, privacy-friendly, and transparent cyber environment provided by the consensus mechanism and cryptographies.

### A. From Code is Law to Code as Law: Blockchain as an Alternative Architecture

As Lawrence Lessig put forward in his book *Code: And Other Laws of Cyberspace*, the regulation of the Internet is not only about legislation but also includes social norms, market forces, and particularly the architecture, through so-called "pathetic dot theory" [23]. In the pre-Internet era, the world was normally regulated and affected by four main factors: legislation enacted by governments to regulate behaviors directly; social norms gradually shaped by the community; market mechanisms guide rational individuals, and architectural elements subtly influence individuals' behavior [23]. This phenomenon has deepened in the Internet age. Online giants, as intermediaries, have directly influenced users' behavior. They can set rules, impose codes of conduct, and even enforce penalties. This is due to the fact that architecture, as a medium that shapes both physical and digital realms, can significantly influence users' behavior and choices by default through the work of coders.

The emergence of blockchain technology presents new opportunities and challenges [24]. On the one hand, the blockchain-based mechanism can increase the legal regulation by default, which means that code can be way to achieve legal objectives. On the other hand, a blockchain-based legal system causes certain problems as traditionally, the law is woven by natural language rather than machine-readable code. Issues therefore arise as to how to translate legal language into machine-readable code, which contains the spirit of the law in a less flexible context[4] [8].

In Lessig's theory, the answer to the machine is the machine. In other words, 'the best way to regulate a code-based system is through code itself' [8]. In the context of a blockchain-based legal system, a new framework of code-based regulation can indeed be set up to regulate people, companies, and machines. Through blockchain technology and relevant smart contracts, legal rules can be interpreted simply as machine-readable rules that can automatically be executed by architecture. Thus, code-based technical rules could achieve a legal function and play the same role as legislation. To this extent, not only can computer code be a complement to law; it can also be another form of law in cyberspace. Filippi and Hassan describe this tendency as from 'code is law'[5] to 'code as law'[6] [5].

### B. Blockchain as the Foundation of Trust

One of the important reasons that individuals respect the law is trust in, and trustworthiness of, the state. However, the ongoing COVID pandemic has again raised a complicated trust dilemma, especially online, between citizens, government, large private entities, and emerging technologies. People always ask similar questions: can we trust the online news regarding the pandemic? Can we trust the man who said he is vaccinated and does not wear a mask? Can we trust the COVID contact tracing app not to leak users' sensitive data to third parties, although the government developed the app? These are all emerging but long-standing issues. Generally, if people want to trust something or someone, there should be multiple trust sources to prove one fact; for example, they can listen to the public service broadcaster, scientists, friends, or online gossip networks for information on vaccines and masks. However, the Internet and digital technologies have broken this balance.

Multiple news sources are emerging every day online, but users cannot easily distinguish the truth. Many transactions are conducted without authenticity; Sztompka and Balázs argue that we normally relied on communal trust, public trust, and private trust to establish the trust before the Internet era. However, these three important trust infrastructures are becoming unreliable in the online environment [25]. For example, the familiarity that arises from ethnic or religious relations, and communities with a shared location or past are becoming unreliable ways to establish communal trust [6]. By the same token, public trust infrastructures such as public education, public service media, and the public press are constantly challenged and undermined by a flow of online key opinion leaders (KOL) and online news platforms. Private trust infrastructures often establish trust through their services or commodities on the market. Independent entities such as banks, insurance companies, and other independent companies establish trust for a service fee. However, with the emergence of the Internet giants, a few oligarchies appeared. Because of their prominence and monopolies, it remains doubtful whether trust can be established[7] [26]. However, blockchain and similar systems create a restrictive and prescriptive technical environment that establishes trust through embedded traceability.

---

[4] A more detailed analysis of machine-readable legislation is provided in the below section.

[5] "Code is Law" is developed from Lessig's theory, which refers to using code to implement specific rules into technology.

[6] "Code as Law" refers to relying on the technology itself both to define and implement mandatory laws.

Unlike traditional centralized control, blockchain allows any user to store a local copy of the blockchain and propose new blocks, laying the foundation for a trustless environment. In contrast, nowadays trust is built in a trusted third party. In the physical world, trust is realized through personal reputation, social norms, or government endorsement, for instance through legislation and contracts [27]. For example, the reason why clients trust their bank is that the bank as a legitimate organization would not find it worthwhile to harm its clients' interests [6]. More importantly, trust is also partially built on previous experience and mostly based on recourses when something bad happens. Such recourses include immediate responses and guidance, regulatory intervention, and the option of private legal action. However, legal compliance generates expensive transaction costs from legal and economic perspectives when both parties are strangers. Similarly, in cyberspace, users need an online intermediary to establish trust to trade. However, centralized architecture causes various issues associated with the online gatekeepers, including data privacy, anti-competition practices, etc. In contrast, blockchain can overturn the traditional architecture and redefine online trust by design. A trustless environment can be established through three overlapping dimensions in the blockchain: users, nodes, and miners.

Because the records on the blockchain cannot be duplicated, manipulated or faked, visibility of (and for) every user is increased. Immutable and timestamped records of transactions endow blockchain with the ability to promote trust based on the 'endorsement of code.' Therefore, blockchain can establish trust without third parties in cyberspace since all transactions are verified and recorded across the whole network.

### C. Blockchain and Transparency in Data Protection

One of the most appealing aspects of blockchain technology is the degree of transparency it can provide. Blockchain is designed as a transparent framework in which anyone can join the network (especially in the public chain), and can therefore can see all information on the chain. As discussed above, the hash-based cryptographic chain makes the chain traceable with appended-only characteristics, and increases overall transparency and accountability.

Therefore, some scholars argue that it is better to describe blockchain as a type of ideology rather than a single technology [9]. Trust is achieved through cryptography rather than institutions or third parties. However, convergence between the law and distributed ledgers is currently lacking. For example, to what extent should the transparency or accountability of these concepts stipulated in a national constitution be translated or transplanted to the context where the law is expressed and enforced through code? By the same token, is the concept of transparency in the technology field equal to the legal one? Therefore, we should scrutinize both law and technology to see if the concept of transparency in law is equal to that in technology, particularly in the blockchain.

Transparency, as required by the GDPR[8] [20], is one of the key elements in protecting users' data privacy rights, as only if the data subject knows who is processing their data and for what purpose can they exercise their data privacy rights. In data protection law, transparency is also the ground of users' rights to information and access to personal data (Art. 13 and Art. 14). Data subjects need such transparency to identify the data controller/processor and exercise their data rights. Therefore, transparency in the GDPR implies that any data related to the data subject should be (i) concise, (ii) easily accessible and easy to understand, (iii) in clear and plain language, and (iv) additionally, where appropriate, visualization can be used [20]. From this perspective, blockchain may struggle to meet its legal obligations.

From another angle, transparency in the GDPR also indicates that the data subject should be aware of certain risks regarding their personal data processing [20]. It connects the principle of transparency to the principle of accountability. It obliges the data controller to demonstrate that personal data is processed transparently. If the data subject anticipates any potential risks, then they can withdraw their consent at any time. According to Article. 29 Working Party (hereinafter: Art. 29 WP), the concept of transparency is quite user-centric rather than legalistic [28]. From this perspective, blockchain-based technologies could assist data protection law to achieve its goal. Since blockchain is designed as a transparent and decentralized mechanism, every user and node can be involved in data transactions. Therefore, data subjects on the chain can see how their personal data are used and processed.

### D. Blockchain and Machine-readable Legislation

The idea behind machine-readable legislation is to replace rules written in a legal language with computer code to achieve legal regulation in cyberspace [29]. Unlike traditional legislation, machine-readable rules can directly alter environmental architectures to enforce the law. Legislators can state exactly which behavior is forbidden or encouraged to shape cyberspace without extra work. If laws need to be modified or adapted to reflect social development, legislators can also adapt legislation quickly through machine-readable laws and implement them rapidly in cyberspace. Meanwhile, because no intermediaries are needed, execution costs will be reduced. The efficiency of execution will also be increased thanks to the use of mathematical consensus algorithms instead of bureaucratic mechanisms.

Represented by smart contracts, blockchain technology has the potential to enable machine-readable laws [30]. Smart contracts work by following simple "if/when… then…" conditional statements that are written into the code on a blockchain. When predetermined conditions have been met and verified, the code will automatically execute the action [8]. In a smart contract, there can be as many provisions as necessary to satisfy the participants' requirements to complete the task. All parties or participants must determine how the transactions and their data will be represented on the blockchain in the "if/when... then…" form to establish the terms and conditions. It is also necessary to explore a few possible exemptions and dispute resolution frameworks. In machine-readable contracts, blockchain offers a convenient portal for all parties signing the contract with an immutable and auditable archive. The sophisticated encrypted identification system can also help the signatory to identify the legitimate parties. By appointing a certain contract towards a specific address of blockchain records, a machine-readable contract can be pointed to the address, and hence be readable when the address is identified.

---

[7] For example, Facebook users may suffer losses due to massive personal data breaches.

[8] Article 5(1), (Regulation (EU) 2016/679 of the European Parliament and of the Council of 27 April 2016 on the Protection of Natural Persons with Regard to the Processing of Personal Data and on the Free Movement of Such Data, and Repealing Directive 95/46/EC, 2016).

There are many potential application scenarios. For instance, it could be used for registration services, ticket purchases, and refunds. The blockchain will be updated to record transactions upon completion. It guarantees the integrity of the transaction, which means that no one can change or modify it. Therefore, if a dispute arises in the future, the record can be directly used as electronic evidence in court, as is discussed further below.

However, blockchain as a form of machine-readable law also poses different challenges. Unlike translation from one language to another language, translating natural legal language into computer code language is not easy. Because of the room for interpretation in legal language, natural legal language usually needs a lawyer, judge, or third party to interpret its application to a real case [29]. Meanwhile, different provisions and articles of the legislation normally follow the principles and aims of different laws. Any overly mechanical interpretation of a particular legal term or provision can lead to misinterpretation of the law by rendering that particular law contrary to the principles and purposes of the original law as a whole [8]. Additionally, as discussed above, before entering a smart contract, every party should list possible exemptions. However, the exceptions cannot be exhausted by enumeration; there will probably be other new exceptions, and machine-readable laws will struggle to recognize the new exceptions. Therefore, a dispute resolution framework is required.

From a more general viewpoint, blockchain-based machine-readable rules may deprive citizens of the right to choose to obey or disobey the law. In economic legal theory, the reason why the law is a deterrent is that there is a cost to breaking it. In other words, when the cost of breaking the law is low, illegal behavior will thrive. However, from another perspective, if the cost of illegality can be tolerated, people may choose to break the law in a specific situation. In the context of machine-readable laws, however, this ultimately changes the architecture of cyberspace, which lacks flexibility and implies that a person can no longer choose to break the law in a particular situation.

To conclude, similar to other technologies, blockchain-based machine-readable laws bring both opportunities and challenges. It is clear that machine-readable laws are easier to execute, and the cost of implementation will be reduced. Meanwhile, once a dispute occurs, the data on the blockchain can be directly used as electronic evidence thanks to the immutability of DLT. The law enforcement process will be more transparent because all data is verified by every relevant party. However, some drawbacks of machine-readable rules are also apparent due to their lack of flexibility. Therefore, further research should focus on exploring the circumstances in which machine-readable laws can be advantageous and those in which such laws do not apply. A case-by-case analysis may be required to explore the function of machine-readable laws in different sectors. For example, in criminal law or the case of extremist crimes, these behaviors have to be prohibited so that they can be forbidden by machine-readable rules through the architecture of cyberspace.

E. Blockchain-based Electronic Evidence: New Understanding of Evidence in Cyberspace

With the burgeoning development of Internet industries, traditional forms of evidence and forensic methods have been challenged, and electronic evidence (e-evidence) is gradually being used and admitted [31]. In order to be legally recognized, e-evidence normally needs to satisfy three conditions: objectivity, relevance, and legality [31]. However, e-evidence is more vulnerable to tampering, deletion, and instability compared to traditional evidence [32]. Thus, most e-evidence is time-sensitive, and without a quick and easy way to lawfully obtain and preserve it, it will be significantly more difficult to recover the evidence.

Currently, there are two main ways to obtain and store e-evidence lawfully. Firstly, the e-evidence is obtained and stored by the individual or representative by taking photos, downloading, or taking screenshots. Secondly, the e-evidence can be obtained through notarization. The notary will certify the authenticity and legality of civil actions, facts, and documents in accordance with the statutory procedures upon application by a natural person, legal person, or other organization. However, there are also a few drawbacks. The first way is easy and inexpensive, but the evidentiary capacity and probative force are weak, and the e-evidence is less likely to be admissible in court. The second way carries a high cost in time and money.

No unified approach currently exists that can be applied across all scenarios and circumstances. Thanks to the features of blockchain, however, the dilemmas of e-evidence's authenticity and integrity can be effectively eliminated [33]. One good example is the emergence of Non-fungible Tokens (NFTs) into the practice of copyright [33] and acceptance as a major reference and copyright certificate for the ownership of the data and content, which effectively prevents hacking and other activities that may destroy evidence [34]. Therefore, blockchain inherently makes e-evidence self-tamper-proof and requiring self-authentication endorsement to ensure its authenticity and integrity [32]. Moreover, blockchain not only serves as a repository for e-evidence, but it can also act as a platform to generate e-evidence, store e-evidence, and notarize e-evidence [35]. Indeed, this combination of a generating, preserving, and notarizing platform is a good reflection of recordism. For example, the Chinese Hangzhou Internet Court has already employed such a system, in the form of a consortium blockchain-based system which connects courts, notarial agencies, forensic institutions, and all users [32]. The application also can be used for daily transaction services for enterprises or individual users. When a dispute occurs, the affected enterprises or individual users can directly use the application to show the e-evidence with a notarial certification, to avoid issues relating to authenticity and integrity.

F. Blockchain's Implications for Data Privacy Law

Blockchain as a new way to store and process data has both positive and negative implications for privacy and data protection. On the one hand, blockchain can achieve legal objectives (especially data protection law) in realizing transparency, privacy protection, a zero-trust mechanism, and decentralization. However, on the other hand, it interferes with the data protection legal regime, because the latter is designed around the concept that there is at least one data controller, and data can be deleted through the data controller. However, under blockchain, it is challenging to comply with these legal principles and rights [36]. This section will focus on what advantages the blockchain could bring to data protection law, and how to mitigate the potential conflicts between blockchain and traditional data protection law.

1. De-identification to Achieve Privacy Preservation

One of the advantages of blockchain is that it can enable trust and auditable transactions through a decentralized network [37]. In identity-sensitive sectors, blockchain can provide a secure and transparent record of who has shared sensitive data with whom, while protecting the content of the data itself. Blockchain is premised on mathematically derived pseudonyms for verification of the distributed ledger. This mechanism can de-identify users' data and achieve privacy protection.

Although the GDPR regards pseudonymous data as personal data since it can be linked to the individual by combination, the design of the blockchain is inherently a privacy-friendly framework because it uses asymmetric encryption as a safeguard to protect users' identification. According to Recital 28, "the application of pseudonymization to personal data can reduce the risks to the data subjects concerned and help controllers and processors to meet their data-protection obligations" [38]. Meanwhile, pseudonymization is also encouraged [38]. The GDPR considers the pseudonymization of data as a means of data protection by design and data protection by default [38], and thus considers it to be a privacy-friendly measurement.

However, blockchain mechanisms also challenge the data protection laws represented by the GDPR. Pseudonymization is easier to achieve than the strict requirement of anonymization, but the law does not provide enough incentives to do so [39]. Data protection law should incentivize the proportional use of pseudonymous data when pseudonymization can provide enough protection to users' privacy in different context-based environments. Because of recent improvements in big data, the dichotomy between anonymous data and pseudonymous data has become blurred and unpractical. In contrast, a dynamic approach to data classification demonstrates flexibility and pragmatism, and is sometimes even more privacy friendly.

2. Blockchain as a Tool to Boost Data Free Flow

One of the legal objectives of the GDPR is to boost personal data free flow. Many jurisdictions also seek to establish digital data marketplaces where personal and non-personal data can be shared, traded, and flow freely. For example, the European Commission considers that the data marketplace can be used to unlock the value of data for the Digital Single Market, and can therefore render the EU more competitive in the context of artificial intelligence [40]. Similarly, Chinese data trade marketplaces are booming. Many data marketplaces, for example the Shanghai Data Exchange Centre [41] and the Guiyang Global Big Data Exchange Centre [42], are taking shape. Data marketplaces powered by blockchain can offer users confidentiality and convenience in data sharing by listing their data on the marketplace. Potential buyers are eligible to use the data with authentication records from data providers when the trade has been secured by the marketplace [36].

Blockchain can offer considerable advantages in building up data trade marketplaces while providing extra data privacy protection. Firstly, as blockchain is a decentralized framework, the data exchange can be operated without a central trusted intermediary. Because blockchain can enable the zero-trust mechanism [21], it can be used in data transactions and to facilitate the free flow of data, as well as allowing participants to provide proof of a statement without disclosing identity information. Furthermore, a blockchain-based data governance tool can establish a smart contract, which can automate data sharing and reduce transaction costs, and it can also be used to achieve legal objectives regarding users' data rights.

3. Blockchain Empowers User with Data Rights

Due to the characteristics of blockchain, it can give users more control over their personal data [18]. As stipulated in Recital 7 of the GDPR, informational self-determination is one of the principles of the GDPR. It implies that data subjects shall have the ability to determine their personal data. Therefore, the GDPR vests various data rights in the data subject; for example, the right of access (Article 15 GDPR) and the right to data portability (Article 20 GDPR) [43]. Blockchain makes it easier and safer to achieve these data rights. On the one hand, it gives users more access to their data in accordance with Article 15 GDPR. Data subjects can easily find out who accessed their personal data, when, and for what purpose, as well as how the personal data is processed [19]. This satisfies the GDPR requirement and provides users with extra accessibility to their data.

Apart from users' right to access, another advantage of blockchain is that it is easier to achieve data portability in practice. The right to data portability is a new type of data right which empowers users to move, copy, or transfer their data from one platform to another [39]. However, data portability is often difficult to achieve due to low interoperability between platforms in practice [20]. Blockchain technology addresses the issue by mitigating transfer costs, as it can facilitate direct relationships and transactions between an individual and multiple organizations. Because the user's data can be cryptographically secured by a private key, the data subject can grant or revoke third-party access to their data any time. If the user wants to switch the data to a new service provider, they only need to grant that provider permission to access the data and revoke the permission from the old one. Therefore, blockchain can make it easier for data subjects to smoothly transfer their data without relying on any organization to port it, guaranteeing data integrity and security during the data transfer.

Furthermore, blockchain can facilitate data subjects in monitoring their data porting process while exercising their right to portability [19]. Because the data subjects may not understand the porting procedure, they face the challenge of identifying lapses and mistakes, and if necessary, taking steps to correct these situations. Blockchain-based data management tools can provide data subjects with oversight of the porting process without the need to impose obligations on platforms.

However, it is worth noting that blockchain and DLT must never accomplish these goals automatically; rather, they must be intentionally designed through interdisciplinary

### G. The Potential Disadvantages of Blockchain in a Legal Sense

Although the sections above have outlined blockchain's various advantages in establishing a legal implementation environment, some potential disadvantages and the premises of these advantages must be acknowledged. The following sections explore the disadvantages and premises from two perspectives. Firstly, the premises of blockchain-enabled features in relation to law will be discussed from a technical perspective. Secondly, blockchain's challenges to existing legislation will be explored.

#### 1. The Premises of Blockchain in Relation to Law

Firstly, the trust achieved by blockchain is not absolute and unconditional, because the consensus algorithm and particular cryptography can be vulnerable over time and suffer from cyberattacks [44]. As a result, the legality and integrity of blockchain-based e-evidence is susceptible to challenge in the court, especially as cryptography faces the risk of cyberattack. Secondly, different types and architectures of blockchain have different levels of credibility, transparency and trustworthiness. For example, private blockchain, consortium blockchain, and public blockchain all have different levels of credibility and transparency, ranging from low to high. It follows that the type of blockchain is a very significant factor in determining the impact and efficacy of the blockchain. Therefore, when blockchain is applied to the judicial system, a detailed case-by-case analysis of its architecture and functionality is required to determine its actual credibility and transparency in the context of the law.

#### 2. Blockchain's Challenges to Existing Legislation

Blockchain, as an innovative methodological cognition, also challenges existing legislation and legal systems. Firstly, it is incompatible with current legislation. For example, blockchain-based transparency is not entirely in line with its legal stipulation. As was mentioned above, transparency in the GDPR requires a concise, easily accessible and understandable (where appropriate, visualization can be used) description to the data subject. From this perspective, blockchain struggles to achieve its legal obligations, because it deploys a decentralized framework in which it is difficult to distinguish the role and responsibilities of data controllers, processors and even users. Furthermore, the sophisticated system could raise the social barriers for data subject to learn and aware such data processing.

Additionally, a blockchain-based e-evidence system also challenges the legal system, because not every jurisdiction currently recognizes the e-evidence generated and stored by blockchain. This situation may require the amendment of current evidence and forensics laws in the EU and China as the law must keep pace with technological developments to ensure that, on the one hand, it can still regulate technology; and on the other hand, that technology can easily achieve legal objectives. Meanwhile, the design stage of blockchain-based evidence systems also requires the combined efforts of technical staff and legal practitioners to ensure that blockchain e-evidence is also fully compliant with the requirements of legal practice.

Lastly, blockchain-based mechanisms are not fully compatible with the current data protection law. As mentioned above, there is no central data controller/processor, and it is difficult to modify or delete data to guarantee data integrity and transparency. However, data protection law (represented by the GDPR) empowers users with the right to erasure and rectification, which is tough to implement. Meanwhile, it is unclear who is responsible for executing these right requests.

#### 3. The Conflicts between Blockchain and Data Protection Law: A Legal Interpretivism Required

As has been discussed, there are several conflicts between blockchain and current data protection law, mainly because the GDPR was designed based on legislators' past experience in cyberspace. However, blockchain inaugurates a new type of data storage which subverts many traditional cyberspace cognitions. Therefore, the data protection law should respond to these challenges to offer legal certainty for the development of emerging technologies. Broad legal interpretivism may be helpful.

In terms of the right to erasure and rectification, although the GDPR stipulates that data subjects shall have the right to ask the controller to erase their personal data, there is no specific definition of the meaning of erasure, rectification, or deletion [39], leaving room for data deletion on blockchain and requiring further legal amendment to explain this issue. By using legal interpretivism and a teleological approach, erasure can not only be explained as the complete erasure or deletion of the data itself but also making data content inaccessible to others and eliminating subsequent impact on the data subject.

Therefore, de-linking to data should also be seen to exercise the right to erasure. In the *Google Spain* case [45], the judge confirmed that de-listing information from research results could be considered to amount to erasure. Although the de-listing was requested by the claimant, who did not ask Google to delete the original data, it is still inspiring for the practical application of the right to erasure in other cases, because in cyberspace, the complete deletion of data is not an easy task to achieve. If nobody can access the data and there is no subsequent impact on the data subject, then that should amount to erasure from a teleological interpretation. Therefore, in the context of blockchain, the right to erasure can also be applied in this way. Although deletion is hard, removing the link and destroying the private key, making data inaccessible, is not difficult. Thus, legal interpretivism and a teleological approach are necessary to explain such issues.

As mentioned above, in the context of DLT, it is difficult to identify the data controller, processor, and user [19]. For instance, nodes in principle only see encrypted and hashed data where data has been modified when put on the blockchain. However, a single node cannot exercise the request of the data subject. By the same token, the definition of 'data controller' and 'processor' must be interpreted teleologically; otherwise, it is impossible to distinguish the responsibility of each participant. According to the explanation of the Article 29 Working Party [46], data controllers can normally determine the purposes and means of processing. This means that the data controller(s) have the ability to respectively determine the 'why' and the 'how' of certain processing activities. Therefore, in a private chain or a consortium chain, the participant who can determine the why and how data is processed is the data controller(s). Similarly, in the context of the public chain, the participant who initiates the chain and can determine the data processing manner satisfies the definition of the data controller, and in some circumstances, miners and nodes can also be joint-data controllers. In this case, miners and nodes must follow the main data controllers' requirement to undertake their own responsibilities.

## VI. IMPLICATIONS OF BLOCKCHAIN FROM A FINANCIAL PERSPECTIVE

Blockchain brings a revolution in information processing, storage and exchange, enabling many new applications in finance, corporate governance, and digital currencies. Researchers believe that the implementation of blockchain could add value to community governance and collaborative production, providing privacy to the identity of participants. During the process, an ordinary participant can become a trader, a miner, a retailer, and a customer simultaneously, breaking down market barriers and reducing the information gap between participants in the community, as all records are clearly visible.

Through tamper-proof records, blockchain-enabled business applications can easily achieve transparency and trust. These new features empower the traditional business model with new meanings and promote the evolution of new business settings. This section analyses the impact of blockchain on financial and other businesses and industries. It specifically examines the positive and negative effects of the blockchain-enabled information revolution on corporate governance and central bank digital currencies.

### A. Blockchain and corporate governance

Smart contract is a vital technology which is able to improve the efficiency of financial markets[9]. Such extensions enable blockchains to keep self-executing transactions. Due to its automaticity and time-saving features, the usage of blockchain could resolve several long-standing issues in the corporate governance area [47]. Moreover, blockchain has the potential to yield unparalleled transparency, for example in identifying the ownership proportion of debt and equity owned by investors or senior management and minimizing the risk of malfeasance on the part of listed companies, exchanges, and regulators. For market participants, blockchain allows faster, more economic acquisitions of shares, but with far less confidentiality compared to centralized system supervision [48]. In the blockchain setting, financial records would become far more transparent, reducing the chance of profit manipulation and other accounting tricks, and related party transactions such as shadow banking would be far more accessible. Even though the user's identity is largely protected, the occurrence of fraud and criminality still cannot be ruled out, through the combination with big data.

The execution of blockchain also boosts the speed and reduces the cost of trading in the financial market. Traditional stock trading in the United States typically takes three business days to settle and transfer ownership from seller to buyer, generating direct and indirect fees for brokers, individuals, and institutional investors. In contrast to traditional transactions, by dropping the existence of the market middleman, the blockchain system enhances market liquidity. Such a characteristic is a critical threshold boosting the funding flow and catering to the interest and demand of fund managers and investors [49]–[52]. Although stock exchanges are still likely to play a role in connecting buyers and sellers, the cost-effectiveness and faster trading time of the blockchain system will likely make it a more attractive option for market attendees.

Besides transparency and low trading cost, blockchain technology has been presented as a possible voting platform for all sorts of elections. Numerous studies [53]–[55] have explored the existing difficulties with corporate elections, including inaccurate voter lists, limited ballot distribution, and even chaotic vote counting. In a blockchain system, the token can be regarded as a valid ticket to record the choices made by voters. As has been claimed[56], [57], voting via blockchain could encourage shareholders to actively engage in corporate governance and vote on a broader range of issues more frequently. In a similar spirit, election fraud, exemplified by the unfounded 2020 US presidential election rumors, can effectively be avoided via a blockchain-based voting system.

### B. Discussion between blockchain and central bank digital currencies (CBDC)

When the blockchain originated in the market, the debate between the central system under supervision and the distributed ledge system generated governmental concerns and wide discussion. It is notable that the emergence of blockchain triggered the innovation of central bank digital currencies. As the bellwether, the leading cryptocurrency Bitcoin is taking over a recognized part of the financial landscape, therefore worldwide central banks are becoming more interested in the potential application of digital money. Fearing the possible risk of losing sovereign currency issuance, central banks in developed countries and unions such as the UK, the US, the EU, New Zealand, and elsewhere are considering establishing their own central bank digital currencies (CBDCs). Along with this, extensive discussion and conceptual work has been published in a flood of white papers recently [44], [58]–[60].

Although CBDC is commonly connected with Distributed Ledger Technology, the most common central bank attitude towards that linkage is disapproval, especially as the centralized payment system is the foundation of their previous absolute control of currency. Despite the concerns raised by the authorities, a blockchain-based ledger offers several potential breakthroughs that can be independently utilized, allowing discussion of the particular features of the blockchain system that can be used in a CBDC setting. The first of these is smart contract, which can be independently implemented over a number of different kinds of ledgers, including a centralized system. Further, security and efficiency targets could be achieved through restricting the broad function accessible inside a smart contract programming language, and cryptography could also be employed to strengthen the CBDC system. Based on one of the white papers [44], it is necessary to upgrade the cryptographic approaches since particular cryptographic functions can be weakened over time and vulnerable to high-turnover cyberattacks. In a nutshell, the innovation of blockchain technology has undoubtedly triggered a revolution in the financial system and boosted the invention of CBDC. Although it is still in the early development phase in the majority of countries, the People's Bank of China established its digital currency research institution in 2016 to explore the elaboration of the blockchain application and efficient market governance. In April 2020, a trial test of Digital Yuan (CBDCs) was executed in relation to the forthcoming Olympic Winter Games and achieved more than US$ 13.8 billion transactions by the end of 2021 [61].

---

[9] Smart contracts are programs recorded on a blockchain that execute when certain circumstances are satisfied.

C. Potential obstacles to the financial industry's move toward a blockchain-based business model

As has been discussed, cryptocurrencies like Bitcoin, which has followed its own unique path to growth and expansion, are vulnerable to cybercrime and illegal use. The emergence of digital assets originated on a global scale, as opposed to a national one, and with individual investors rather than institutional participants, raising irrational herd behavior and the free rider issue. Since cryptocurrencies are decentralized, it is challenging to determine the sponsoring organization. Even if the sponsoring institution can be determined, the issuer or platform may be located outside of certain jurisdictions, making legal enforcement difficult [43], [62], [63]. The elevated probability of cybercrime relative to traditional financial markets is a result of the centralized form of crypto exchanges and the hot wallet effect. If hackers are successful in stealing coins from cryptocurrency exchanges and moving the assets to their own personal wallets, then millions of wallet addresses can be automatically generated via blockchain to transfer Bitcoin [62]. These stolen funds could have taken a mysterious path involving several different token addresses on various blockchains, and hackers are also able to thwart authorities' efforts to track and stop money laundering.

Therefore, in the context of the existing challenge of cybercrime, the Deloitte 2021 blockchain survey recognized that the most significant challenges to the adoption of digital assets relate to cybersecurity concerns and regulatory constraints [63]. The US and other regulators produced a hodgepodge of guidelines after the advent of Bitcoin in 2008 and the launch of the Bitcoin network in January 2009 to accommodate the rapid speed of crypto and blockchain innovation. For instance, in the United States, it is still unclear whether regulators such as the SEC (Securities and Exchange Commission), the Federal Reserve and the Commodity Future Trading Commission will collaborate, or whether Congress will set up a relevant regulatory framework. We anticipate that in the near future, the US authorities will provide more guidelines addressing issues associated with cybersecurity, AML, securities registration, anti-fraud, tax, and transaction reporting hazards in connection to cryptocurrencies. Other economies have advanced this process with more transparent frameworks, like the EU's Markets in Crypto-Assets framework which claimed that investors would enjoy some protection and that reserves are required by stable coin issuers to prevent large withdrawals, as happened in a local region [63].

Besides the widespread illegal activities taking place in the digital market and the lack of regulation protecting retail investors, the second concern coming from the financial institution pioneers is the underdevelopment of the blockchain-based financial infrastructure [63]. A private chain is more likely to be the predominant model chosen by companies instead of a public chain, and therefore the primary concern becomes how to exchange information with a trustworthy third party. The advantages extend well beyond financial sectors, particularly in cases where numerous businesses access and share the same data and need insight into transaction records. This is often a costly, wasteful, and untrustworthy procedure. Due to technological barriers and privacy concerns, third-party data access is usually restricted. Private and permissioned DLT platforms allow companies to communicate and share data in a secure manner, guaranteeing that only the necessary degrees of data access are granted to third parties that have been specifically authenticated and deemed trustworthy. Data privacy is a growing concern if businesses seek to exchange data across corporate and sector borders in order to boost cooperation and trust among ecosystem partners without compromising data integrity or privacy.

VII. BLOCKCHAIN'S POTENTIAL FROM A TECHNOLOGICAL PERSPECTIVE

Through its decentralization of organizations and finance, blockchain is not only having a significant impact on society and the finance industry, but is also bringing innovation to our daily lives. Additionally, blockchain demonstrates great potential in combination with other technologies, by advancing an innovative methodology. In other words, the benefits of blockchain are not limited to the development of other technologies, but the development of other technologies can also promote the growth of blockchain. This section analyzes how blockchain can be harnessed more sustainably by integrating it with other techniques, and how technological improvements can help to address the challenges raised by blockchain applications.

A. Distribution system of income and wealth

Blockchain in the global economy has been widely accepted in the form of cryptocurrencies. However, blockchain reaches far beyond the scope of cryptocurrencies. It has also reshaped the world with its encrypted nature and opened up the opportunity for every individual with the power to do so, to mine or mint tokens that can be used in every possible way [64]. Tokens are the essential carrier of the value of blockchain. It makes sense when people pass on tokens to perform transactions on the blockchain. Doing so brings a breakthrough in the system of income distribution, where the individual has total control of their earnings and earns extra fees when they are serving the network. Since the distribution of tokens is direct and transparent, it equalizes participants' roles and creates the legal person's unique jurisdiction in the blockchain aspect, where a human being is mapped with an encrypted identity used to represent the legal person. To the greatest extent possible, a widely open distributed ledger provides the ultimate audit to the economic system [64].

B. Functional Consensus for sustainable development

Consensus, the step of getting agreement among all peer nodes, should never be a weakness of blockchain, but instead should be a way of promoting it. However, ever since the rise of Bitcoin, blockchain has been linked with high power consumption, low efficiency, and resource waste, with an associated environmental impact. In the first quarter of 2021, Cambridge estimated that consumption to be 84.6 TWh, and it expanded the estimation rapidly later to a staggering 112.82 TWh of electricity each year, which is a huge amount of electricity and thus carries a carbon footprint [65]. A common criticism is that the consensus used by blockchains is not productive but a puzzle-solving game with no real value in reality. In the near future, Proof-of-Work consensuses will remain the safest and most accepted public chain solution, and it will be hard to replace them without substantial changes to the current acknowledgment of Bitcoin and the transitional understanding to greener consensus. On the other hand, there are practical functional consensuses, which utilize the puzzles or the consensus itself as an essential service; for instance, the Proof-of-Replication (PoRep) [13] used by Filecoin makes use of the replication status of file sharing for miner election [66], hence providing a data query service in parallel with the PoRep consensus. Functional consensus is a major step forward in building a sustainable ecosystem for blockchain. It is also challenging to design a functional consensus while considering the efficiency, availability, security, and integrity of the network.

An application-specific integrated circuit (ASIC) is a type of computing machine that has been designed to solve particular problems; in our case, the search for hash. It produces enormous heat while running, and the cooling of them costs extra electricity via air conditioners, which further increase the temperature around the ASICs. Room heating using mining ASIC has been well-practiced in homes with a low electricity price and a low required temperature [67]. The energy-hungry machine produces roughly 2 kW of heat with a radiator design fitted to the cooling unit. The heating cost savings during the winter and the added value from mined crypto assets reach balances in a short cycle and eventually cover the cost of the ASIC hardware. This has been recognized as a mitigation to the environmental impacts of ASIC mining activities, and the market for ASIC radiator rental has grown, with more and more interest from greenhouses and winter gardens.

### C. Guidelines for prohibiting non-renewable energies and carbon tax for cryptos

Considering huge carbon and greenhouse gases emissions and the ongoing waste of non-renewable energies, non-renewable energies should not be used to power blockchain networks, especially the cryptocurrency network, in line with the commitments made by the jurisdictional governing body and local communities. The legislation should therefore consider guidelines and standardizing the energy source for blockchain application, reflecting its role in, and responsibility to minimize, environmental impacts. Essential services powered by blockchain should be allowed to use any energy sources as per the local jurisdiction and regulations. However, Proof-of-Work based cryptos, for instance, should refrain from using coal/oil/gas-generated electricity and pay a carbon tax based on power rating. By setting up guidelines and regulations for blockchain energy policies, the environmental impact of blockchain can be greatly reduced and a renewable future-proof legal status provided for blockchain maintenance. Compared to the difficulty of taxing cryptocurrencies, taxing the mining machine would help to reduce greenhouse gas emissions and allow climate benefit via good use of the tax collected [68].

### D. Storage revolution for blockchain in the future

Blockchain provides immutable records to the public with its persistent storage of all records. It would, however, be a great waste of resources if the records were allowed to keep growing to infinity. There is some mitigation available in reducing the storage size on each node, that comes at a great risk to data integrity and availability and tempers the blockchain's basic principle of immutability. Hence, the need for persistent distributed storage of all records in a distributed way is a critical issue for future blockchain storage. There are as yet no answers on how to persistently store all records in a distributed manner at an acceptable cost, but there are candidates for this feature.

With the exploration of new materials, biological genetic materials, i.e., DNA and RNA, are good examples of storing a full copy of the record in a distributed way. In other words, any individual DNA carries the same piece of information from its parent DNA when it performs mitosis during cell division. The double helix shape of the linkage also resembles the immutability of blockchain, considering its hash linkage. Also, the shape of folded protein is a way to efficiently store the copy of the blockchain in the near future, thanks to its compact size and persistent nature. Recent protein-based storage [69] progress provides a practical way to distribute and store the same piece of information with a highly stable shape and link. The protein can be rapidly scanned to find the differences in their structure and hence figure out the data changes. It is a suitable medium for long-term blockchain storage and the fast validation of data records in response to the various governance, law, and financial challenges raised in this paper.

### E. Endogenous and resilient security infrastructure

In a blockchain-backed society, cyber security remains a critical challenge for the fully network-dependent blockchain infrastructure. Blockchain offers ubiquitous security to all participants via public key-based identification and authentication. While the blockchain continues to offer reliable records of legal and financial activities, it can also be considered as a security infrastructure [22], [70], [71], which is also a perfectly resilient work in nature. Through the usage of blockchain-based security infrastructure, all entities are forced to adopt Public-key-as-Identity, thus offering a ubiquitous authentication capability to all without third-party involvement. By having the Public-key pairs, the owner is able to mutually authenticate each of them by signing the blockchain wallet address while claiming and proving ownership of the blockchain wallet address.

## VIII. CONCLUSION

Developments in sciences and technology will drive the progress and transformation of society, which could have a great societal impact, and vice versa. The contribution of this paper is that it examines blockchain's implications for society in detail, from methodologies, to legal systems and financial markets. It finds that the principle and design philosophy of blockchain still has great potential, despite its negative impact on existing law and finance. Regarding methodological cognition, the blockchain-driven information revolution refreshes the cognition of the source of power and provides a new bottom-up architecture for governance. This new cyberspace architecture helps to achieve some legal objectives, such as trust and transparency, while challenging the existing legal regimes. Therefore, a legal interpretivist approach is required to resolve the conflict between blockchain and current legislation.

Furthermore, this paper finds that blockchain could enable new applications in corporate governance by breaking down market barriers and reducing the information gap between participants. However, blockchain also brings new risks to the financial market, particularly from cybercrime and illegal use, which need be addressed. It is not enough to take a purely technological perspective, as more interdisciplinary solutions are required. Future research could focus on collecting and analyzing empirical evidence of blockchain's implications for society.